# Magnetic domain-wall motion by propagating spin waves

Dong-Soo Han, Sang-Koog Kim,[a] Jun-Young Lee

*Research Center for Spin Dynamics & Spin-Wave Devices, Seoul National University Nanospinics Laboratory, Department of Materials Science and Engineering, Seoul National University, Seoul 151-744, Republic of Korea*

Sebastian J. Hermsdoerfer, Helmut Schultheiss, Britta Leven, Burkard Hillebrands

*Fachbereich Physik and Forschungszentrum OPTIMAS, Technische Universität Kaiserslautern, 67663 Kaiserslautern, Germany*

We found by micromagnetic simulations that the motion of a transverse wall (TW) type domain wall in magnetic thin-film nanostripes can be manipulated via interaction with spin waves (SWs) propagating through the TW. The velocity of the TW motion can be controlled by changes of the frequency and amplitude of the propagating SWs. Moreover, the TW motion is efficiently driven by specific SW frequencies that coincide with the resonant frequencies of the local modes existing inside the TW structure. The use of propagating SWs, whose frequencies are tuned to those of the intrinsic TW modes, is an alternative approach for controlling TW motion in nanostripes.

[a] Author to whom all correspondence should be addressed, e-mail: sangkoog@snu.ac.kr

Electric-current driven manipulation of domain wall (DW) motion in patterned magnetic thin-film nanowires is rapidly increasing in interest in the areas of magnetism and spintronics devices,[1-6] owing to its potential applications to solid-state data-storage and data-processing devices, such as race-track memory[7] and logic gates.[8] Recently, fundamental understandings of the nontrivial dynamics of DW motion driven by applied magnetic fields and/or spin-polarized electric currents have been obtained from several experimental,[1-6,9,10] numerical simulation,[11-14] and theoretical[15-17] studies. Meanwhile, only a few studies have been conducted on the influence of a DW subject to a flow of spin waves (SWs) passing through it from the aspect of a manipulation of SWs by the static structure of DWs.[18,19]

Although numerous researches on electric-current and/or magnetic field driven DW motion have been carried out, SW driven DW motion has not been reported so far. In this letter, we report on the results of a study on transverse wall (TW)-type DW motion driven by SWs passing through it via interaction between them. Thus, we propose that traveling SWs are an alternative means for the manipulation of DW motion in nanostripes.

In this study, we chose an approach of micromagnetic numerical calculations on a model system of a Permalloy (Py) nanostripe of $t =10$ nm thickness, $l = 3005$ nm length, and $w$ = 50 nm width, where a head-to-head TW-type DW was placed at the center position ($x$ = 0), as shown in Fig. 1. This TW structure has its characteristic potential hill along the longitudinal direction caused by a finite length of the nanostripe, as seen in the inset of Fig. 1. For the simulations, we selected a cell size of $5 \times 5 \times 10$ nm$^3$ and material parameters corresponding to Py: the saturation magnetization $M_s$ = $860 \times 10^3$ A/m, the exchange stiffness $A_{ex}$ = $1.3 \times 10^{11}$ J/m, and the Gilbert damping constant $\alpha$ = 0.01. We used the object-oriented micromagnetic framework (OOMMF) code[20] that utilizes the Landau-Lifshitz-Gilbert equation of **M** dynamic motion.[21] For the generation and injection of single, monochromatic, plane-wave-like spin waves (SWs), we applied a single harmonic sinusoidal field $\mathbf{H} = H_0 \sin(2\pi\nu_H t)\hat{\mathbf{y}}$ along the $y$ axis with the field frequency $\nu_H$ and field amplitude $H_0$ only to a local area ($5 \times 50 \times 10$ nm$^3$) at the left-end edge.[22]

The resultant motion of the TW induced by the SW flow are shown in Fig. 2, which displays top-view, serial snapshot images of the spatial distribution of the local normalized z component of the magnetization $M_z/M_s$ for the TW motion driven by specific frequencies, here $f_{SW}$ = 18 and 13 GHz, generated with $\nu_H$ =18 and 13 GHz, respectively, and with $H_0$ = 1.0 kOe.

It is evident that the motion of the TW is driven by SWs of a specific frequency of $f_{SW}$ = 18 GHz via interaction between the TW and the SW flow, but the TW does not move for $f_{SW}$ =13 GHz. To obtain more quantitative information on the relationship between $f_{SW}$ and TW motion during a time duration, $\Delta t$ = 25 ns, the average velocities $\bar{\upsilon}$ is plotted versus $f_{SW}$ in Fig. 3(a) for a frequency range of 10 to 45 GHz in steps of 0.5 GHz. Note that SWs having $f_{SW}$ < 12 GHz are not allowed to propagate through the narrow nanostripe ($w$ = 50 nm) due to the lateral width confinement[23], for which the potential barrier[24] is ~ 12 GHz. As seen in Fig. 3(a), three strong and two weak peaks appear, such that $\bar{\upsilon}$ =1.1, 5.9, 4.6, 2.1, and 0.8 m/s at $f_{SW}$ =14.5, 18, 24, 27, and 32 GHz, respectively. This $\bar{\upsilon}$-versus-$f_{SW}$ curve reveals that there exist specific frequencies that drive efficiently TW motion with their own corresponding velocities. The dependence of $\bar{\upsilon}$ on the amplitude (or intensity) of the propagating SWs (i.e., $H_0$ in its linear response regime) is also shown in Fig. 3(b) for a specific frequency of $f_{SW}$ = 18 GHz with $H_0$= 0.1, 0.15, 1.0, 4.0, 5.0, and 8.0 kOe. Below $H_0$=0.15 kOe the TW does not move, but for the case of $H_0 \geq$ 0.15 kOe, the value of $\bar{\upsilon}$ increases with the magnitude of $H_0$. It is worthwhile to note that the critical $H_0$ value (or threshold SW intensity) for the TW motion in the nanostriple ($w$= 50 nm) is ~ 0.15 k Oe for 18 GHz.

The longitudinal displacements of the TW motion driven by the five peak values of $f_{sw}$ and with an amplitude of $H_0 = 1.0$ kOe are also shown in Fig. 3(c), indicating that those TW motion are almost steady in the observation range of 0-25 ns. In contrast, for longer times the displacement is rapidly increasing when the TW reaches the right-end edge, showing that the TW motion accelerates when it is getting close to the edge. This is due to the potential hill of the TW with respect to the long axis, being caused by the finite size of the long axis of the nanostripe, as shown in the inset of Fig. 1. The average velocity of TW motion caused by the potential hill is estimated to be 0.17 m/s from the simulation result on a relaxation of the TW positioned at the off-center position, $x = +10$ nm. Therefore, the linear displacements (constant velocities) at the initial motion shown in Fig. 3(c) are associated with compensatory effects of the decrease in the intensity of SWs acting on the DW being moved further to the right-end edge, and the increase in the velocity of the TW driven by its potential hill.

To elucidate the underlying mechanism of the pronounced sensitivity of the TW motion to several specific frequencies of SWs, as shown in Fig. 3(a), we conducted additionally micromagnetic simulations on the same model as that shown in Fig. 1, but with the injection of SWs of frequencies of 0 to 45 GHz, by using a sinc-function field $\mathbf{H} = H_0 \sin(2\pi\nu_H t)/(2\pi\nu_H t)\hat{\mathbf{y}}$, with $H_0 = 10$ Oe, and $\nu_H = 45$ GHz. Figure 4 shows the comparison between the dispersion relations ($f_{SW}$ versus $k_x$) obtained from FFT (Fast Fourier Transformation) of the temporal evolution of the SW propagation at a fixed position of $y$ = 25 nm in nanostripes with and without TW.[24] In both dispersion curves, there is no spectral combination in the region of $f_{SW}$ < 12 GHz, which is due to the potential barrier of SWs caused by the stripe-width confinement, as already mentioned. There is similarity in both dispersions. In the region of $k_x$ > 0, there are two curves with positive curvature in the lower and higher branches. The former corresponds to the forward propagation of SWs along the $x$ direction and the latter includes the contribution of SWs propagating along the transverse direction.[24] Also, there are distinct differences between the two dispersions. For the nanostripe with a TW [Fig. 4(a)], a strong signal appears in the branch for $k_x$ <0, which is absent in the nanostripe without TW. This is due to reflection of SWs by the TW. In the case of the presence of TW, additionally three local modes appear at frequencies, 18, 24, and 27 GHz, caused by the presence of the TW structure in the nanostripe. These three characteristic local modes coincide with the frequencies of the three strong peaks found in the $\bar{\upsilon}$-versus-$f_{SW}$ curve shown in Fig. 2(a).

To further investigate this close relationship between the $\bar{\upsilon}$ sensitivity to $f_{SW}$ and the local modes inside the TW, we plotted FFT power versus $f_{SW}$ in Fig. 5, which was extracted from the dispersion relation shown in Fig. 3(a), by conducting the FFT of the local $M_z/M_s$ distribution at the position of $y = +25$ nm along the longitudinal direction in the limited region of $x = -45$ nm to $+45$ nm containing the TW. In the comparison with the $\bar{\upsilon}$-versus-$f_{SW}$ curve, the indicated five peak positions agree very well between both curves, although their intensities are only in qualitative agreements.

To conclude, TW motions can be resonantly excited by interaction of SWs with the internal modes of the TW. The TW velocity varies with the frequency and amplitude of the propagating SWs. These results provide an alternative means for controlling DW motion in nanostripes.

This work was supported by Creative Research Initiatives (ReC-SDSW) of MEST/KOSEF. Financial support by the DFG within the Priority Programme SPP1133 "Ultrafast magnetization processes" is gratefully acknowledged.

# Reference

[1] M. Hayashi, L. Thomas, R. Moriya, C. Rettner, and S.S.P. Parkin, Science **320**, 209 (2008).

[2] M. Laufenberg, W. Bührer, D. Bedau, P.-E. Melchy, M. Kläui, L. Vila, G. Faini, C. A. F. Vaz, J. A. C. Bland, and U. Rüdiger, Phys. Rev. Lett. **97**, 046602 (2006).

[3] A. Yamaguchi, T. Ono, and S. Nasu, K. Miyake, K. Mibu, T. Shinjo, Phys. Rev. Lett. **92**, 077205 (2004).

[4] M. Kläui, P.-O. Jubert, R. Allenspach, A. Bischof, J. A. C. Bland, G. Faini, U. Rüdiger, C. A. F. Vaz, L. Vila, and C. Vouille, Phys. Rev. Lett. **95**, 026601 (2005).

[5] M. Tsoi, R. E. Fontana, and S. S. P. Parkin, Appl. Phys. Lett. **83**, 2617 (2003).

[6] J. Grollier, P. Boulenc, V. Cros, A. Hamzić, A. Vaurès, and A. Fert, Appl. Phys. Lett. **83**, 509 (2003).

[7] S. S. P. Parkin, M. Hayashi, and L. Thomas, Science **320**, 190 (2008).

[8] D. A. Allwood, G. Xiong, C. C. Faulkner, D. Atkinson, D. Petit, and R. P. Cowburn, Science **309**, 1688 (2005).

[9] Y. Nakatani,, A. Thiaville, and J. Miltat, Nat. Mat. **2**, 521 (2003).

[10] M. Hayashi, L. Thomas, Ya. B. Bazaliy, C. Rettner, R. Moriya, X. Jiang, and S. S. P. Parkin, Phys. Rev. Lett. **96**, 197207 (2006).

[11] A. Thiaville, Y. Nakatani, J. Miltat, and N. Vernier, J. Appl. Phys. **95**, 7049 (2004).

[12] Andrew Kunz, J. Appl. Phys. **99**, 08G107 (2006).

[13] J.-Y. Lee, K.-S. Lee, S. Choi, K. Y. Guslienko, and S.-K. Kim, Phys. Rev. B **76**, 184408 (2007).

[14] S.-K. Kim, J.-Y. Lee, Y.-S. Choi, K. Y. Guslienko, and K.-S. Lee, Appl. Phys. Lett. **93**, 052503 (2008).

[15]G. Tatara and H. Kohno, Phys. Rev. Lett. **92**, 086601 (2004).

[16]Z. Li and S. Zhang, Phys. Rev. B **70**, 024417 (2004).

[17]K. Y. Guslienko, J.-Y. Lee, and S.-K. Kim, IEEE Trans. Mag. **44**, 3079 (2008).

[18] R. Hertel, W. Wulfhekel, and J. Kirschner, Phys. Rev. Lett. **93**, 257202 (2004).

[19]C. Bayer, H. Schultheis, B. Hillebrands, R. L. Stamps, IEEE Trans. Mag. **41**, 3094 (2005).

[20]See http://math.nist.gov/oommf.

[21] L. D. Landau and E. M. Lifshitz, Phys. Z. Sowjetunion **8**, 153 (1935); T. L. Gilbert, Phys. Rev. **100**, 1243 (1955).

[22]K.-S. Lee, S.-K. Kim, J. Appl. Phys. **104**, 053909 (2008)

[23]K.Y. Guslienko and A. N. Slavin, J. Appl. Phys. **87**, 6337 (2000); K. Y. Guslienko, R. W. Chantrell, and A. N. Slavin, Phys. Rev. B **68**, 024422 (2003).

[24]S. Choi, K.-S. Lee, K. Y. Guslienko, and S.-K. Kim, Phys. Rev. Lett. **98**, 087205 (2007).

**Figure captions**

Fig. 1. (color online) Model Py nanostripe of rectangular cross section with indicated dimensions and a transverse wall (TW) type DW positioned at the center, $x = 0$. The in-plane **M** direction is represented by the color-coded direction wheel, along with the streamlines with small arrows. The black box represents the region where plane-wave-like spin waves (SW) are excited. The inset shows the potential hill of the TW along the longitudinal direction ($x$-axis), which was obtained from the calculation of the total energy as a function of the $x$ position of the DW displacement.

Fig. 2. (color online) Top-view snapshot images of the spatial distribution of the in-plane orientations of local $\mathbf{M}_s$, displaying the temporal evolution of the motion of the TW, driven by propagating SWs of $f_{SW} = 18$ and 13 GHz in (a) and (b), respectively. The vertical white-dotted line denotes the center position, $x = 0$.

Fig. 3. (color online) (a) Average velocity of TW versus the frequency of driving SWs calculated in the range of 10 to 45 GHz. Local peaks are marked by different symbols at 14.5, 18.0, 24.0, 27.0, and 32.0 GHz. The red dashed horizontal line corresponds to the velocity of TW (~ 0.17 m/s), caused by the potential hill of the TW without the injection of SWs. (b) Average velocity of the TW driven by $f_{SW} = 18$ GHz and with different values of $H_0 = 0.1, 1.0, 4.0, 5.0,$ and 8.0 kOe. (c) Longitudinal (x-axis) displacements of the TW subject to a flow of SWs with frequencies according to the peaks in Fig. 3(a) in a field of $H_0 = 1.0$ kOe. (d) Displacement of the TW for $f_{SW} = 18$ GHz with $H_0 = 1.0$ kOe for sufficiently long time for the TW to reach the right-end edge. The red dashed

line corresponds to the right end edge of the nanostripe.

Fig. 4. (color online) Dispersion relations of SWs propagating through the nanostripe with a TW in (a) and without it in (b). For the case of (b), the structure was saturated in the +x direction.

Fig. 5. (color online) Comparison of the FFT power versus $f_{SW}$ (blue solid line), representing the local modes inherent in the TW, and the average velocity of the TW movement versus $f_{SW}$ (red solid line), shown in Fig. 3(a) .

Fig 1.

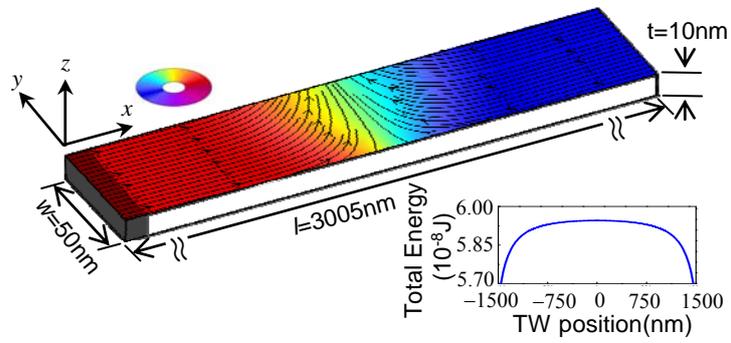

Fig 2.

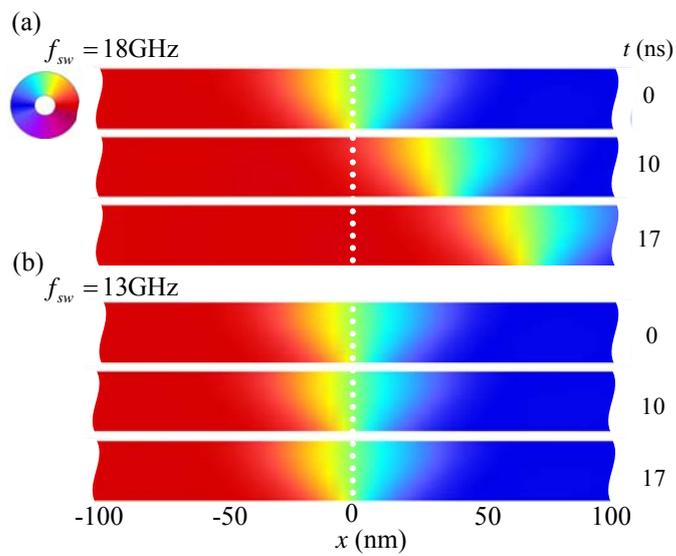

Fig 3.

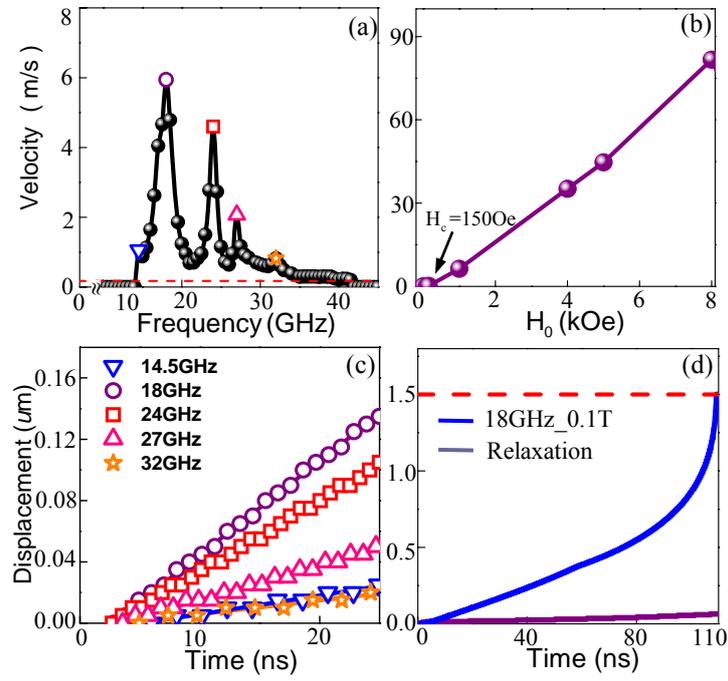

Fig 4.

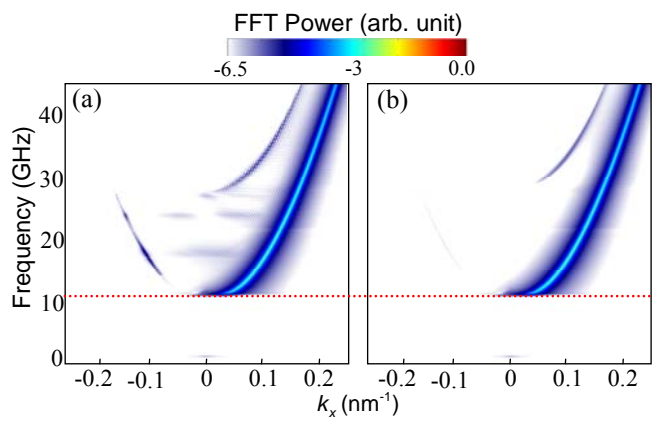

Fig 5.

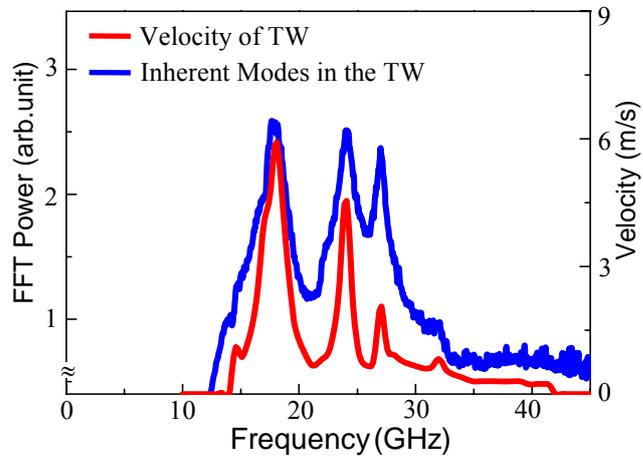